\documentclass[showpacs,aps,prb,twocolumn]{revtex4}
\usepackage[english]{babel}
\usepackage{epsfig}
\usepackage{graphicx}
\usepackage{amsmath}
\usepackage{amssymb}
\begin{document}

\title{A model for the phase separation controlled by
doping and the internal chemical pressure in different cuprate
superconductors}

\author{K.~I.~Kugel} \affiliation{Institute for Theoretical and
Applied Electrodynamics, Russian Academy of Sciences, Izhorskaya
Str. 13, Moscow, 125412 Russia} \affiliation{Department of
Physics, Loughborough University, Loughborough LE11 3TU, United
Kingdom}

\author{A.~L.~Rakhmanov} \affiliation{Institute for Theoretical
and Applied Electrodynamics, Russian Academy of Sciences,
Izhorskaya Str. 13, Moscow, 125412 Russia} \affiliation{Department
of Physics, Loughborough University, Loughborough LE11 3TU, United
Kingdom}

\author{A.~O.~Sboychakov} \affiliation{Institute for Theoretical
and Applied Electrodynamics, Russian Academy of Sciences,
Izhorskaya Str. 13, Moscow, 125412 Russia}

\author{Nicola Poccia} \affiliation{Department of Physics,
University of Rome ``La Sapienza'', P.le A. Moro 2, 00185 Rome,
Italy}

\author{Antonio Bianconi} \affiliation{Department of Physics,
University of Rome ``La Sapienza'', P.le A. Moro 2, 00185 Rome,
Italy}

\begin{abstract} In the framework of a two-band model, we study
the phase separation regime of different kinds of strongly
correlated charge carriers as a function of the energy splitting
between the two sets of bands. The narrow (wide) band simulates
the more localized (more delocalized) type of charge carriers. By
assuming that the internal chemical pressure on the CuO$_2$ layer
due to interlayer mismatch controls the energy splitting between
the two sets of states, the theoretical predictions are able to
reproduce the regime of phase separation at doping higher than 1/8
in the experimental pressure-doping-$T_c$ phase diagram of
cuprates at large microstrain as it appears in overoxygenated
La$_2$CuO$_4$. \end{abstract}



\date{\today}

\maketitle

\section{Introduction}\label{intro}

The mechanism driving the emergence of a quantum macroscopic
coherent phase that is able to resist to the de-coherence effects
of high temperature remains a major topic of research in condensed
matter. The realization of this macroscopic quantum phase in doped
cuprates close to the Mott insulator regime has stimulated a large
amount of investigations on the physics of strongly correlated
metals. Most of theoretical papers treated models of a homogeneous
system made of a single electronic band (or models of multiple
hybridized bands reduced to a single effective band), with a large
Hubbard repulsion.

There is growing agreement that the solution of the problem  of
high-$T_c$ superconductivity  requires the correct description of
the normal state where spin, charge, orbital and lattice degree of
freedoms compete and the functional phase emerges in a complex
system with two main components showing mesoscopic phase
separation. Here we consider a theoretical model of the mesoscopic
phase separation in a two-band scenario of two strongly correlated
electronic fluids. This simple model grabs the key physics of the
anomalous normal phase in cuprates exhibiting the phase separation
as a function of charge density and the energy splitting between
the two bands. This allows the understanding of the different
superconducting phases in different cuprate families, i.e., the
new 3D phase diagram where the critical temperature depends on the
doping and pressure~\cite{38}. The motivation of this theoretical
work is based on the results of recent experiments using angular
resolved photoemission spectroscopy (ARPES) \cite{1,2,3,4,5} and
scanning tunnelling spectroscopy (STM) \cite{6,7} providing
compelling experimental evidence for dual nature of charge
carriers and the nanoscale phase separation of the two components
in two different spatial domains in cuprate high-$T_c$
superconductors \cite{16,15,45}.

A clear case for the phase separation of the two types of charge
carriers is overoxygenated La$_2$CuO$_{4+y}$ where the
interstitial oxygen ions are mobile above 180 K and stimulate the
phase separation of the two different kinds of dopant holes
\cite{44,19,40,41,20,42,18,43,21}. Currently, from the analysis of
magnetic neutron scattering experiments, there is an agreement for
the frustrated mesoscopic phase separation at doping larger than
1/8 in Sr-doped La214, Y123, and Bi2212 between a first more
delocalized component that does not show spin fluctuations and a
second more localized electronic component, showing stripe-like
spin fluctuations \cite{22,23}. Several reviews and books have
been published on the two-component scenario and phase separation
in cuprates \cite{18,43,21,24,25,26,27}. Here we focus not on the
well studied phase separation in the underdoped regime, near the
Mott phase, between a hole-poor antiferromagnetic  phase and a
metallic hole-rich phase but on the phase separation in the
overdoped regime \cite{44,19,40,41,20,42,18,43,21,22,23} between a
hole poor phase with doping close to 1/8 and a hole-rich phase
with doping close to 1/4.

This phase-separation scenario has been described in
Refs.~\onlinecite{18,22,23} for LaSrCuO and overoxygenated
La$_2$CuO$_4$. It is based on the experimental fact that variation
of the magnetic incommensurability due to spin stripes saturates
at doping 1/8 see Fig. 15 of Ref.~\onlinecite{22}. The residual
magnetic scattering at high doping suggests that one of the phases
has stripe correlations similar to the $x =1/8$  phase, with the
volume fraction of this phase decreasing with $x$, as indicated in
Fig. 17 of Ref.~\onlinecite{22}.  The other phase is presumably
uniformly doped. The picture, then, is that as one increases the
doping beyond $x=1/8$, it becomes unfavorable to accommodate the
additional holes in stripes; instead, patches of the
uniformly-doped phase grow at the expense of the stripe phase. The
maximum $T_c$ seems to occur in a mixed phase region dominated by
the stripe phase.

A similar scenario is now well accepted for understanding the
physics of phase separation in manganites~\cite{dagbook,28,mango}.
It was shown that even in the absence of any specific order
parameter, the presence of two strongly correlated electron bands
leads to the possibility of a phase-separated state \cite{33}.

In Ref.~\onlinecite{33}, the evolution of phase separation was
studied as a function of doping. However, a large amount of data
clearly indicates that the phase separation regime is not only a
function of doping but also of the anisotropic chemical pressure
acting on the CuO$_2$ layers, due to interlayer mismatch
\cite{29,30,31,32}. The chemical pressure is a well established
physical variable that controls the physical properties of
perovskites and it is usually measured by the average ionic radius
of the cations in the intercalated layers or the tolerance factor
$t$, in fact, the internal chemical pressure in perovskites can be
defined as  $\eta = 1-t$. In all perovkites and particularly in
manganites, it is well established that the phase diagram of the
electronic phases depends on the two variables, charge density and
chemical pressure \cite{Dag}. Since the early years of high-$T_c$
superconductivity research the mismatch chemical pressure has been
considered as a key variable controlling the electronic properties
of cuprates only on one family, La214 \cite{32}, however it was
not possible to extend this idea to other families for the
presence of a plurality of intercalated layers with cations having
largely different coordination numbers. Therefore, it was not
possible to compare the average ionic size $\langle r_A\rangle$ in
the intercalated layers and to get the tolerance factor $t$ for
all cuprate families . This problem was solved by obtaining the
internal chemical pressure from the measure of the compressive
microstrain $\varepsilon  = (R_0 -r)/r$ in the CuO$_2$ plane (that
has the same absolute value as the tensile microstrain in the
intercalated layers) where $r$ is the average Cu-O distance and
$R_0=0.197$~nm is the unrelaxed Cu-O distance \cite{29,30,31}.
Therefore, the chemical pressure is proportional to microstrain,
$\eta = 2\varepsilon$.

In this new 3D phase diagram, the phase separation for the
overdoped regime in overoxygenated La214  occurs in a family with
high chemical pressure close to $\eta = 8\%$, while it becomes a
frustrated phase separation in the LaSrCuO, Bi2212, and Y123 that
are in the range of chemical pressure $7\%>\eta >4\%$, while for
cuprates with lower microstrain only very fast critical
fluctuations could be present \cite{29,30,31}. In this paper, we
propose a model of a two-component system made of two different
strongly correlated electron bands where the chemical pressure
controls the energy splitting between the two bands. The phase
separation in the overdoped regime can exist for specific values
of the ratio between the bandwith of the two bands. The critical
point for the transition from a frustrated phase separation  to a
non frustrated phase separation can be obtained by tuning the long
range $1/r$ Coulomb repulsion that frustrates the phase separation
as going from Sr doped to oxygen doped La124.

\section{The model}\label{model}

The existence of the two types of the strongly correlated charge
carriers in cuprates can be described in terms of the two-band
Hubbard model. The Hamiltonian of such a system can be written as
\cite{33}
\begin{eqnarray}\label{H}
H\!&=&\!-\!\!\!\sum_{\langle\mathbf{nm}\rangle\alpha,\sigma}\!\!t_{\alpha}a^
{\dag}_{\mathbf{n}\alpha\sigma}a_{\mathbf{m}\alpha\sigma}%
-\Delta E\sum_{\mathbf{n}\sigma}n_{\mathbf{n}b\sigma}
-\mu\sum_{\mathbf{n}\alpha,\sigma}n_{\mathbf{n}\alpha\sigma}\nonumber
\\
&+&\frac{1}{2}\sum_{\mathbf{n}\alpha,\sigma}U^{\alpha}n_{\mathbf{n}
\alpha\sigma}n_{\mathbf{n}\alpha\bar{\sigma}}%
+\frac{U'}{2}\sum_{\mathbf{n}\alpha,\sigma\sigma'}
n_{\mathbf{n}\alpha\sigma}n_{\mathbf{n}\bar{\alpha}\sigma'}\,.
\end{eqnarray}
Here, $a^{\dag}_{\mathbf{n}\alpha\sigma}$ and
$a_{\mathbf{n}\alpha\sigma}$ are the creation and annihilation
operators for electrons corresponding to bands $\alpha=\{a,\,b\}$
at site $\mathbf{n}$ with spin projection $\sigma$, and
$n_{\mathbf{n}\alpha\sigma}=a^{\dag}_{\mathbf{n}
\alpha\sigma}a_{\mathbf{n}\alpha\sigma}$. The symbol
$\langle\dots\rangle$ denotes the summation over the
nearest-neighbor sites. The first term in the right-hand side of
Eq.~\eqref{H} corresponds to the kinetic energy of the conduction
electrons in bands $a$ and $b$ with the hopping integrals
$t_a>t_b$. In our model, we ignore the interband hopping. The
second term describes the shift $\Delta E$ of the center of band
$b$ with respect to the center of band $a$ ($\Delta E>0$ if the
center of band $b$ is below the center of band $a$). The last two
terms describe the on-site Coulomb repulsion of two electrons
either in the same state (with the Coulomb energy $U^{\alpha}$) or
in the different states ($U'$). The bar above $\alpha$ or $\sigma$
denotes {\it not} $\alpha$ or {\it not} $\sigma$, respectively.
The assumption of the strong electron correlations means that the
Coulomb interaction is large, that is, $U^{\alpha},\,U'\gg
t_{\alpha},\,\Delta E$. The total number $n$ of electrons per site
is a sum of electrons in the $a$ and $b$ states, $n=n_a+n_b$, and
$\mu$ is the chemical potential. Below, we consider the case
$n\leq 1$ relevant to cuprates.

Model~\eqref{H} predicts a tendency to the phase separation in a
certain range of parameters, in particular, in the case when the
hopping integrals for $a$ and $b$ bands differ significantly ($t_a
> t_b$)~\cite{33}. This tendency results from the effect of strong
correlations giving rise to dependence of the width of one band on
the filling of another band. In the absence of the electron
correlations ($n \ll 1$), the half-width $w_a = zt_a$ of $a$ band
is larger than $w_b = zt_b$ ($z$ is the number of the nearest
neighbors of the copper ion). Due to the electron correlations,
the relative width of $a$ and $b$ bands can vary
significantly~\cite{33}.

In the limit of strong correlations, $U^{\alpha}, U'\rightarrow
\infty$, we can describe the evolution of the band structure with
the change of $n$ and $\Delta E$ following the method presented in
Ref.~\onlinecite{33}. We introduce one-particle Green function
\begin{equation}\label{Grf1}
G_{\alpha\sigma}(\mathbf{n}-\mathbf{n}_0,\,t-t_0)=
-i\langle\hat{T}a_{\mathbf{n}\alpha\sigma}(t)a^{\dag}_{\mathbf{n}_0
\alpha\sigma}(t_0)\rangle,
\end{equation}
where $\hat{T}$ is the time-ordering operator. The equations of
motion for one-particle Green function with Hamiltonian \eqref{H}
includes two-particle Green functions
\begin{equation}\nonumber
{\cal{G}}_{\alpha\sigma,\beta\sigma'}(\mathbf{n}-\mathbf{n}_0,\,t-t_0)=%
-i\langle\hat{T}a_{\mathbf{n}\alpha\sigma}(t)n_{\mathbf{n}\beta\sigma'}(t)%
a^{\dag}_{\mathbf{n}_0\alpha\sigma}(t_0)\rangle\,.
\end{equation}

In the considered limit of strong on-site Coulomb repulsion, the
presence of two electrons at the same site is unfavorable, and the
two-particle Green function is of the order of $1/U$, where $U\sim
U_{\alpha},U'$. The equation of motion for
${\cal{G}}_{\alpha\sigma,\beta\sigma'}$ includes the
three-particle terms coming from the commutator of
$a_{\mathbf{n}\alpha\sigma}(t)$ with the $U$ terms of
Hamiltonian~\eqref{H}, which are of the order of $1/U^2$ and so
on. In these equations, following the Hubbard I
approach~\cite{Hub}, we neglect the terms of the order of $1/U^2$
and make the following replacement
$\langle\hat{T}a_{\mathbf{n}+\mathbf{m}\alpha\sigma}(t)n_{\mathbf{n}
\beta\sigma'}(t)
a^{\dag}_{\mathbf{n}_0\alpha\sigma}(t_0)\rangle\to\langle
n_{\mathbf{n}\beta\sigma'}\rangle\langle\hat{T}
a_{\mathbf{n}+\mathbf{m}\alpha\sigma}(t)a^{\dag}_{\mathbf{n}_0
\alpha\sigma}(t_0)\rangle$. As a result, we derive a closed system
for the one- and two-particle Green functions~\cite{33,Hub}. This
system can be solved in a conventional manner by passing from the
time-space $(t,\mathbf{r})$ to the frequency-momentum
$(\omega,\mathbf{k})$ representation. In the case of
superconducting cuprates the total number of electrons per site
does not exceed unity, $n\leq1$. The upper Hubbard sub-bands are
empty, and we can proceed to the limit
$U_{\alpha},\,U'\rightarrow\infty$. In this case, the one-particle
Green function $G_{\alpha\sigma}$ is independent of $U$ and can be
written in the frequency-momentum representation
as~\cite{mango,33} \begin{equation}\label{Ginf}
G_{\alpha\sigma}(\mathbf{k},\omega)=
\frac{g_{\alpha\sigma}}{\omega+\mu+\Delta
E^{\alpha}-g_{\alpha\sigma}w_{\alpha}\zeta(\mathbf{k})}\,,
\end{equation} where $\Delta E^{\alpha}=0$ for $\alpha=a$ and
$\Delta E^{\alpha}=\Delta E$ for $\alpha=b$,
\begin{equation}\label{g}
g_{\alpha\sigma}=1-\sum_{\sigma'}n_{\bar{\alpha}\sigma'}
-n_{\alpha\bar{\sigma}}\,, \end{equation}
$n_{\alpha\sigma}=\langle n_{\mathbf{n}\alpha\sigma}\rangle$ is
the average number of electron per site in the state
$(\alpha,\sigma)$, and $\zeta(\mathbf{k})$ is the spectral
function depending on the lattice symmetry. Since the results does
not vary crucially with the change of the lattice
symmetry~\cite{33}, here we consider the case of the simple cubic
lattice, when $\zeta({\bf k})=-\left[\cos (k^1d)
+\cos(k^2d)+\cos(k^3d)\right]/3$, $d$ is the lattice constant. In
the main approximation in $1/U$, the magnetic ordering does not
appear and we can assume that $
n_{\alpha\uparrow}=n_{\alpha\downarrow}\equiv n_{\alpha}/2$.

Equations \eqref{Ginf} and \eqref{g} demonstrate that the filling
of band $a$ depends on the filling of band $b$ and \textit{vice
versa}. Indeed, using the expression for the density of states
$\rho_{\alpha}(E)=-\pi^{-1} \textmd{Im}\int
G_{\alpha}(\mathbf{k},E+i0)d^3{\bf k}/(2\pi)^3$, we get the
following expression for the numbers of electrons in bands $a$ and
$b$ ($\alpha=a,b$)
\begin{equation}\label{nalpha}
n_{\alpha}=2g_{\alpha}n_0\left(\frac{\mu+\Delta
E^{\alpha}}{g_{\alpha}w_{\alpha}}\right)\,
\end{equation}
where
\begin{equation}\label{n0}
n_0(\mu')=\int\limits_{-1}^{\mu'}dE'\,\rho_0(E')\,,
\end{equation}
and $\rho_0(E')=\displaystyle\int
d^3\mathbf{k}\,\delta(E'-\zeta(\mathbf{k}))/(2\pi)^3$ is the
density of states for free electrons (with the energy normalized
by unity, $|E|\leq1$). The chemical potential $\mu$ in
Eq.~\eqref{nalpha} can be found from the equality $n=n_a+n_b$.

%%%%%%%%%%%%%%%%%%%%%%%%%%% Fig. 1 %%%%%%%%%%%%%%%%%%%%%%%%%%%%
\begin{figure}[htb] \centering
\includegraphics[width=0.5\textwidth]{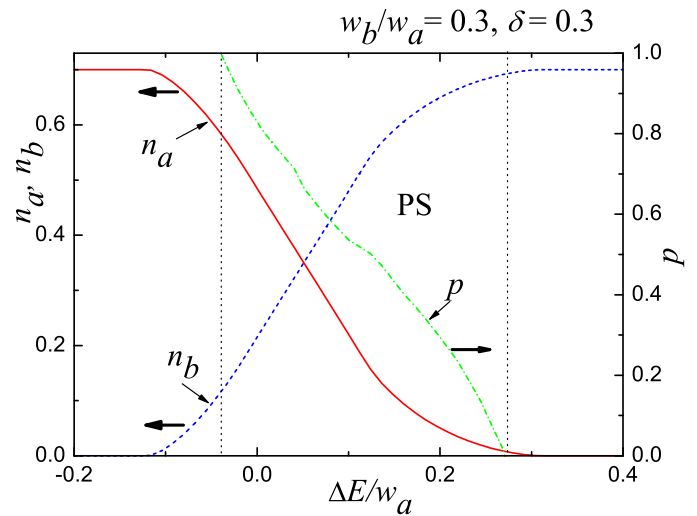}\caption{(Color
online) Evolution of the occupation numbers $n_a$ and $n_b$ of the
bands $a$ and $b$ at fixed doping $\delta =1-n=0.3$ in the absence
of phase separation. The region of phase separation lies between
two vertical dotted lines. There we have two phases: $P_a$,
including mostly $a$ charge carriers and $P_b$ with dominant $b$
carriers. The content of different types of carriers in $P_a$ and
$P_b$ is given by the intersections of $n_a$ and $n_b$ curves with
left and right dashed vertical curves, respectively. The change in
concentration $p$ of phase $P_a$ in the phase-separation region is
shown by the (green) dot-dashed line.}
 \label{Fig1} \end{figure}

 %%%%%%%%%%%%%%%%%%%%%%%%%% Fig. 2 %%%%%%%%%%%%%%%%%%%%%%%%%%%%
\begin{figure}[htb] \centering
\includegraphics[width=0.5\textwidth]{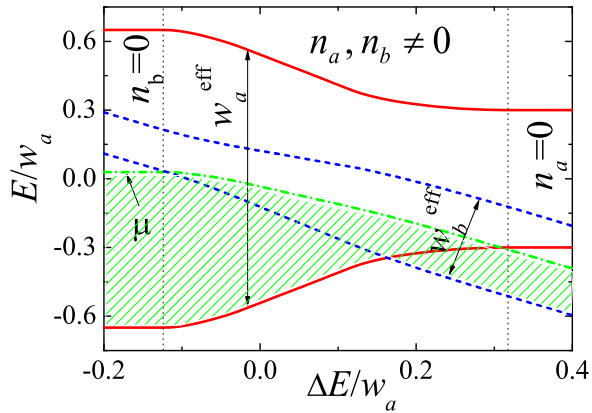} \caption{(Color
online) Effective widths $w_{a,b}^{\rm eff}$ of the $a$ and $b$
bands versus band shift $\Delta E$ at fixed doping $\delta
=1-n=0.3$. The (green) dot-dashed curve illustrates the behavior
of the chemical potential $\mu$; the (green) hatched area under
this curve corresponds to the states occupied by charge carriers.}
\label{Fig2} \end{figure}

When the energy band $b$ is far above the center of band $a$
($\Delta E<0$), there exist only $a$ electrons. With the increase
of $\Delta E$, the chemical potential reaches the bottom of the
$b$ band $-\Delta E-w_b$. At higher $\Delta E$, the $b$ electrons
appear in the system, and the effective width of $a$ band, $w^{\rm
eff}_a=2w_ag_{a}$, starts to decrease. At large positive values of
$\Delta E$, the $a$ carriers in the system disappear and there
exist only $b$ electrons. The plots of $n_a$, $n_b$, and the
effective bandwidth as functions of $\Delta E$ are shown in
Figs.~\ref{Fig1} and \ref{Fig2}, respectively.

%%%%%%%%%%%%%%%%%%%%%%%%%% Fig. 3 %%%%%%%%%%%%%%%%%%%%%%%%%%%%
\begin{figure} \begin{center}
\epsfig{file=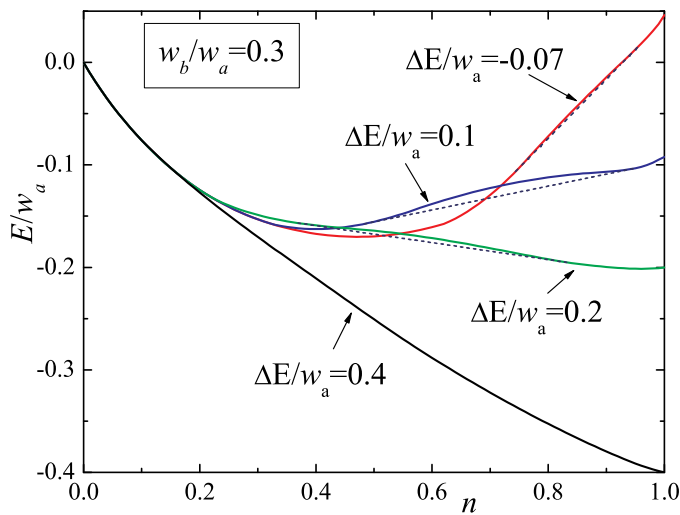,width=0.5\textwidth} \end{center}
\caption{\label{Fig3} (Color online) The energy of the system {\it
vs} doping level $n$ at different values of $\Delta E$. Solid
curves correspond to the homogeneous state, whereas the dashed
curves are the energies of the phase-separated state without
taking into account electrostatic and surface contributions to the
total energy.} \end{figure}

The energy of the system in homogeneous state, $E_{\text{hom}}$,
is the sum of electron energies in all filled bands. Similarly to
Eq.~\eqref{nalpha}, we can write $E_{\text{hom}}$ in the form
\begin{equation}\label{Ehom}
E_{\text{hom}}=2\sum_{\alpha}g_{\alpha}^2w_{\alpha}%
\varepsilon_0\left(\frac{\mu+\Delta E^{\alpha}}{g_{\alpha}
w_{\alpha}}\right)%
-\Delta E\, n_b\,, \end{equation} where
\begin{equation}\label{eps}
\varepsilon_0(\mu')=\int\limits_{-1}^{\mu'}dE'E'\,\rho_0(E')\,.
\end{equation} The dependence of $E_{\text{hom}}(n)$ is shown in
Fig.~\ref{Fig3} at different values of $\Delta E$. We see that
within a certain $n$ range, the system can have a negative
compressibility, $\partial^2E_{\text{hom}}/\partial n^2<0$, which
means a possibility for the charge carriers to form two phases
with different electron concentrations~\cite{29,33}. The negative
compressibility disappears when the centers of the bands are far
apart from each other (see, e.g., the curve corresponding to
$\Delta E/w_a=0.4$ in Fig.~\ref{Fig3}).

The phase separation may be hindered by increase of the total
energy due to surface effects and a charge redistribution.
However, at first we do not take into account this effects. We
consider two phases, $P_a$ (low carrier density) and $P_b$ (high
carrier density), with the number of electrons per site $n_1$ and
$n_2$, respectively. A fraction $p$ of the system volume is
occupied by the phase $P_a$ and $1-p$ is a fraction of the phase
$P_b$. We seek a minimum of the system energy
\begin{equation}\label{PsE0}
E_{\text{ps}}^0(n_1,n_2)=pE_{\text{hom}}(n_1)+(1-p)E_{\text{hom}}(n_2)
\end{equation} under the condition of the charge carrier
conservation $n=pn_1+(1-p)n_2$. The results of calculations of the
system energy in the phase-separated state are shown in
Fig.~\ref{Fig3} by the dashed lines. We see that the phase
separation exists in the range of $n$ where both types of charge
carriers coexist in the homogeneous state. The ratio of the
numbers of $a$ and $b$ carriers is different in different phases.
In the first phase $P_a$, almost all charge carriers are in the
band $a$, while in the second phase $P_b$ the situation is
opposite.

The redistribution of charge carriers in the phase-separated state
gives rise  to the additional electrostatic contribution, $E_C$,
to the total energy. This term in the Wigner-Seitz approximation
was calculated in Ref.~\onlinecite{33}. At $p<0.5$ it can be
written as
$E_{\text{C}}=V\left(n_1-n_2\right)^2\left(R_s/d\right)^2u(p)$,
where \begin{equation}\label{Ec} u(p)=2\pi
p\left(2-3p^{1/3}+p\right)/5, \end{equation} and $V$ is the
characteristic energy of the intersite Coulomb interaction and
$R_s$ is the radius of the spherical droplet of the phase $P_a$
surrounded by the shell of the phase $P_b$. In the case $p>0.5$,
we should replace $n_1\leftrightarrow n_2$ and $p\leftrightarrow
1-p$. The second contribution to the total energy, depending of
the size of inhomogeneities, is related to the surface between two
phases. The corresponding energy per unit volume can be presented
in the form $E_{\text{S}}=pS\sigma(n_1,n_2)/V_0$, where $p<0.5$,
$S$ is the surface and $V_0$ is the volume of the inhomogeneity,
and $\sigma(n_1,n_2)$ is the surface tension. For e spherical
droplets, we have $E_{\text{S}}=3p\sigma(n_1,n_2)d/R_s$. If
$p>0.5$, we should replace $p\to1-p$. Minimization of the sum
$E_{\text{CS}}=E_\text{C}+E_\text{S}$ with respect to $R_s$ allows
us to calculate this value. In doing so, we get at $p<0.5$
\begin{equation}\label{Rs} R_s=d\left(\frac{3p\sigma(n_1,n_2)}{2
V_0(n_2-n_1)^2u(p)}\right)^{1/3}\!\!\!\!\!\!. \end{equation} The
total energy of the inhomogeneous state then reads
\begin{equation}\label{PsE}
E_{\text{ps}}=pE_{\text{hom}}(n_1)+(1-p)E_{\text{hom}}(n_2)+E_{\text{CS}}(R_s)\,,
\end{equation} where $R_s$ is given by Eq.~\eqref{Rs}. The surface
energy comes from the size quantization and it was estimated in
Ref.~\onlinecite{33}. The electrostatic and surface contributions
to the energy related to an inhomogeneous charge distribution
reduce the range of $n$, in which the phase separation is
favorable, see the phase diagram in Fig.~\ref{Fig4}.

\section{Results and discussion}\label{discus}

\begin{figure}[htb] \centering
\includegraphics[width=0.5\textwidth]{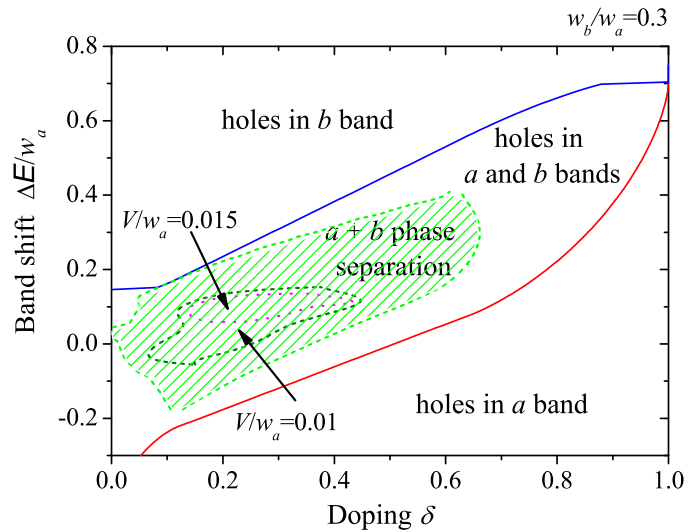}\caption{(Color
online) The phase diagram of model \eqref{H} in the (doping, band
shift) plane at the ratio of band widths $w_b/w_a = 0.3$. Below
the lower (red) solid line, there are charge carriers only of $a$
type, whereas above the upper (blue) curve - only $b$ carriers.
Between these lines, there appears the region of phase separation
(marked by (green) hatching). The charge disproportionalization in
the phase-separated state can substantially reduce this region:
the arrows indicate the phase separation regions at $V/w_a = 0.01$
and 0.015, where $V$ is the characteristic energy of intersite
Coulomb interaction.} \label{Fig4} \end{figure}

The undoped state of the cuprates corresponds to one electron per
site ($n = 1$) in the model used in Ref.~\onlinecite{33}. The
number of itinerant holes $\delta$ is related to $n$ as $\delta =
1 - n$. In general, the relationship between $n$ and $\delta$
could be more complicated~\cite{34}, however, for the present
considerations such corrections are not of principal importance.
The phase diagram of the model~\eqref{H} in the $(\delta, \Delta
E)$ plane is drawn in Fig.~\ref{Fig4}.  In this figure, below the
lower (red) solid line,  we have the charge carriers only of $a$
type, whereas above the upper (blue) curve, there are only $b$
carriers. If we ignore the possibility of the phase separation,
the relative number of $a$ and $b$ charge carriers varies
gradually between these two lines. The evolution of the occupation
numbers $n_a$ and $n_b$ of the two bands with $\Delta E$ at a
fixed doping is illustrated in Fig.~\ref{Fig1}. Note also that the
effective widths of the $a$ and $b$ bands also vary with the band
shift due to the electron correlation effects. This is illustrated
in Fig.~\ref{Fig2}. Taking into account possible phase-separated
states results in a significant modification of the phase diagram
in the range of intermediate doping. In the hatched (green) region
in Fig.~\ref{Fig4} the homogeneous state becomes unfavorable and
the system separates into two phases ($P_a$ and $P_b$) with
different numbers of charge carriers per site $n_1 \approx n_a$
and $n_2 \approx n_b$. The electrostatic contribution to the
energy related to an inhomogeneous charge distribution reduces the
doping range, in which the phase separation is
favorable~\cite{33}. In Fig.~\ref{Fig4}, we illustrate that a
relatively small energy loss due to the charge
disproportionalization leads to a substantial decrease in the area
of the phase-separation region: compare the areas indicated by
arrows corresponding to $V/w_a = 0.01$ and 0.015 ($V$ is the
characteristic energy of intersite Coulomb
interaction~\cite{Shenoy}) and the whole hatched area
corresponding to $V=0$. Note that at low hole doping ($n$ close to
one), the antiferromagnetic (AF) correlations are dominant, which
requires a special analysis.

Estimating the contribution of the long-range Coulomb interaction
to the energy of the phase-separated state, we assume the simplest
droplet-like geometry of the inhomogeneities. The phase separation
occurs in the range of parameters where the energy of homogeneous
state as a function of doping has a negative curvature
corresponding to the negative compressibility~\cite{mango,33}. It
was widely discussed in literature that long-range Coulomb
interaction in the systems with negative compressibility can give
rise to more complicated geometry of the phase separation
(stripes, layers, rods, etc.), see Refs.~\onlinecite{low,ort} and
references therein. Thus, a due account of the long-range Coulomb
interaction could reproduce different superstructures (stripes, in
particular) observed in the cuprate superconductors near the
optimum doping. However, the proper analysis of the inhomogeneity
geometry requires a further study based on a more complicated
model.

%%%%%%%%%%%%%%%%%%%%%%%%%%% Fig. 4 %%%%%%%%%%%%%%%%%%%%%%%%%%%%

Now let us discuss the relation of the above model to the
experimental situation in the copper-based perovskites, where two
types of charge carriers and inhomogeneous (phase-separated) state
are observed. The inhomogeneous state in cuprates corresponds to
the coexistence of two phases. One of them is characterized by  a
superstructure (charge ordering, stripes, etc.) and another one
has no superstructure. The state with charge (or spin)
superstructure corresponds to a higher degree of localization and,
therefore, to a smaller value of the hopping integral. Naturally,
a charge carrier may hop either retaining short-range order and
gaining in the potential energy or hop in an arbitrary way with
larger hopping integral thus gaining in the kinetic energy. The
former corresponds to our $b$ state and the latter to the $a$
state. In our analysis, we did not consider any ordering, which
arises in the next-order approximations. In particular, magnetic
order requires taking into account the terms of the order of
$t^2_{a,b}/U$ and the charge ordering implies allowing for the
Coulomb interaction of carriers at different sites
(nearest-neighbor at least).

The relative position of the two bands, $\Delta E$, and hopping
integrals, $t_{a,b}$, depends, in particular, on the chemical
pressure proportional to a microstrain $\varepsilon$ in the
crystal lattice. To describe the experimental phase diagram of
cuprates in the $(\delta, \varepsilon)$ plane~\cite{29,30,31,38},
we should know the relationship between the model parameters and
the chemical pressure. It is natural to assume that the two bands
in the cuprate crystal originate from the double degenerate $e_g$
hole level of Cu$^{2+}$ (configuration $d^9$) in the crystal field
of cubic symmetry. The splitting of this level occurs due to
lattice distortions related to the Jahn-Teller effect, lowering
the cubic symmetry. The chemical pressure distorts the crystal
lattice even more and should affect the value of $\Delta E$
significantly. It is natural to assume that $\Delta E$ and
$\varepsilon$ are linearly related to each other, if
$|\varepsilon|\ll 1$. So, we can write
\begin{equation}\label{Delta}
\Delta E(\varepsilon)=\Delta E(0)+\Delta E_1f(\varepsilon),
\end{equation}
where $f(\varepsilon)$ is a dimensionless function and
$f(\varepsilon)\approx\varepsilon$ at $|\varepsilon|\ll 1$.
Cu$^{2+}$ is a typical Jahn-Teller ion and we can assume that
$\Delta E_1$ is of the order of the characteristic Jahn-Teller
energy, which is larger than $t_a$ (see, e.g.,
Refs.~\onlinecite{33,Shenoy,35} and references therein). The
effect of the microstrain on the relative band positions can be
significant since the value $\Delta E$ arises due to splitting of
the originally degenerate levels. In the same time, small strains
give rise only to a small corrections to the bandwidth. So, the
ratio $t_b/t_a$ is considered further on as independent of
$\varepsilon$. Note also that the values of the intersite Coulomb
interaction $V$ characteristic of perovskites is of the order of
$0.1-0.01t_a$ (see, e.g., Ref.~\onlinecite{Shenoy} and references
therein). Thus, the value $V=0.015zt_a$ used below is quite
reasonable.

\begin{figure}[htb] \centering
\includegraphics[width=0.5\textwidth]{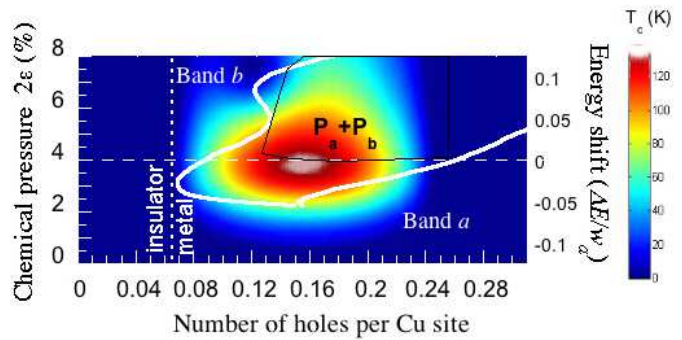} \caption{(Color
online) The values of superconducting transition temperature $T_c$
from 0 for dark blue to 135 K for dark red is shown in a color
plot as function of chemical pressure (2$\varepsilon$) and doping
(holes number per Cu site). The white curve corresponds to the
phase separation region given by the two-band Hubbard model
\eqref{H} corresponding to $w_b/w_a=0.3$ and $V/w_a=0.01$ (see
Fig.~\ref{Fig4}). Phases $P_a$ and $P_b$ include mostly the
carriers of $a$ and $b$ types, respectively. The black solid line
is the boundary of the phase-separated state deduced from neutron
scattering and anomalous diffraction experiments for cuprates
(La214, Bi2212, and Y123 systems).} \label{Fig5} \end{figure}

Bearing this in mind, we compare the theoretical phase diagram in
Fig.~\ref{Fig4} with the experimental 3D phase diagram of
cuprates~\cite{29,38} in Fig.~\ref{Fig5}. The left-hand $y$ scale
is the mismatch chemical pressure $\eta$ related to microstrain as
$\eta = 2\varepsilon$ and the $x$ axis is the doping (the number
of holes per Cu site). The color plot represents the values of
critical temperature in different superconducting cuprate
families. The plot shows the fit of the experimental data of a
large number of materials with the convolution of a parabolic
curve with the maximum at $T_{\rm max}$ for $T_c$ as function of
doping, and an asymmetric Lorentzian for $T_{\rm max}$ as a
function of the mismatch chemical pressure with the maximum of 135
K at $2\varepsilon=4 \%$. The $y$ axis in the right-hand side of
the figure gives the energy distance $\Delta E$ between the center
of band $a$ and band $b$ normalized to the width of band $w_a$ of
the more itinerant carriers. The phase diagram involving the
superconducting critical temperature, chemical pressure, and
doping reaches the $T_c$ maximum at $2\varepsilon = 4\%$ and 0.16
holes per Cu sites. Based on the aforementioned consideration, we
can take $\Delta E(0)=-0.133w_a$ and $\Delta E(0)=6.67w_a$ in
Eq.~\eqref{Delta}. We can identify a low-doping insulating phase,
for any chemical pressure, at doping smaller than 0.06, where the
vertical dashed line indicates the line of metal-insulator
transition. The experimental investigations of the novel 3D phase
diagram of cuprates indicate that the homogeneous metallic phase,
with more delocalized states, occurs for both high doping and  low
chemical pressure i.e., in the low right corner of the figure. In
this region, we have in the theoretical model the charge carriers
only in the band $a$. On the contrary, the homogeneous phase made
of localized states where the striped phase appears, occurs at the
corner on the top-left side of the figure. In the theoretical
model, we have in such a region, the charge carriers only in the
band $b$. The superconducting phase occurs in the intermediate
region between these two limiting cases. The phase separation
region predicted by our model (inside the area bounded by the
solid white line) corresponds to the superconducting phase. It is
in the qualitative agreement with the STM, extended X-ray
absorption fine structure (EXAFS), neutron pair distribution
functions (PDF) experiments showing that high-$T_c$
superconductivity occurs in a regime of mesoscopic phase
separation. The present results show that the maximum critical
temperature occurs where the energy splitting between the more
itinerant, band $a$ and the more localized, band $b$ is close to
zero.

%%%%%%%%%%%%%%%%%%%%%%%%%%% Fig. 5 %%%%%%%%%%%%%%%%%%%%%%%%%%%%

The theoretical phase diagram reproduces qualitatively the
experimental results on the phase separation in cuprates (La214,
Bi2212, and Y123 systems) near optimum doping obtained by neutron
scattering and anomalous diffraction techniques~\cite{29,30,31}.
The phase separation arises in the intermediate doping range and
disappears at low and high doping levels. The phase with ``more
itinerant'' electrons exists at small microstrains, ``more
localized'' (and more ordered) phase arises at higher
microstrains, and the phase-separated state is located in the
intermediate range of $\varepsilon$. However, our calculations
predict the phase separation in a broader doping range than in the
experiments. It seems to be a consequence of simplifications used
in the formulation and approximate analysis of the two-band
Hubbard model~\eqref{H}. To improve the agreement with the
experiment it is necessary to take into account specific features
of the lattice and electron structure of the cuprate
superconductors. In particular, we disregard the interband
electron transitions, that is, we neglect the terms
$t_{ab}a^{\dag}_{\mathbf{n} a \sigma}a_{\mathbf{m} b \sigma}$ in
Hamiltonian~\eqref{H} assuming that $t_{ab} = 0$. The doping range
where the phase separation can exist reduces with the increase of
$t_{ab}$~\cite{35}.

As it follows from Fig.~\ref{Fig5}, the superconducting transition
temperature $T_c$ is the highest for the parameters range where
the system is in the phase-separated state. This is an indication
that the mechanism of the phase separation is intimately related
to the phenomenon of the superconductivity. It is worth to note
that in the case when interband coupling $t_{ab}$ is in the range
$t_a < t_{ab} < t_b$, the electron density of states has a peak
near the Fermi level in the parameter range corresponding to the
phase-separated state, where $T_c$ is maximum~\cite{35}. We can
not claim whether this fact is accidental or not.

\section{Conclusions}

Up to now, most of the attention both of experimentalists and
theorists has been addressed to the phase separation in the
underdoped regime between a first undoped antiferromagnetic  phase
and a second doped metallic phase of cuprates. Now, we have an
evidence for mesoscopic phase separation in the overdoped region
of cuprate superconductors where a striped phase at doping 1/8
coexist with a metallic phase with doping close to 1/4. In our
paper, we were dealing just with this situation.

We have presented an emerging theoretical scenario that relating
the phase separation to the chemical pressure. This scenario grabs
key physical aspects of the 3D phase diagram of cuprates. It was
shown that the two-band model is appropriate for the normal phase
of all cuprate superconducting families, where the energy
splitting between the two bands is controlled by mismatch chemical
pressure. In the regime where the two bands are close in energy,
the system is unstable toward the phase separation. The highest
critical temperature of the superconducting transition in cuprates
is attained within the phase-separated state.

\section*{Acknowledgments}

The work was supported by the European project CoMePhS (contract
NNP4-CT-2005-517039), the International Science and Technology
Center (grant G1335), and by the Russian Foundation for Basic
Research, grants 08-02-00212 and 06-02-16691. A.\,O.\,S. also
acknowledges a support from the Russian Science Support
Foundation.

\end{document}